# Environmentally controlled curvature of single collagen proteins

Naghmeh Rezaei, Aaron Lyons and Nancy R. Forde*

Department of Physics

Simon Fraser University

8888 University Drive

Burnaby, BC  V5A 1S6

CANADA

*Corresponding author.  778-782-3161.  nforde@sfu.ca.  ORCID: 0000-0002-5479-7073

**ABSTRACT**

The predominant structural protein in vertebrates is collagen, which plays a key role in extracellular matrix and connective tissue mechanics. Despite its prevalence and physical importance in biology, the mechanical properties of molecular collagen are far from established. The flexibility of its triple helix is unresolved, with descriptions from different experimental techniques ranging from flexible to semirigid. Furthermore, it is unknown how collagen type (homo- vs. heterotrimeric) and source (tissue-derived vs. recombinant) influence flexibility. Using SmarTrace, a chain tracing algorithm we devised, we performed statistical analysis of collagen conformations collected with atomic force microscopy (AFM) to determine the protein's mechanical properties. Our results show that types I, II and III collagens – the key fibrillar varieties – exhibit molecular flexibilities that are very similar. However, collagen conformations are strongly modulated by salt, transitioning from compact to extended as KCl concentration increases, in both neutral and acidic pH. While analysis with a standard worm-like chain model suggests that the persistence length of collagen can attain almost any value within the literature range, closer inspection reveals that this modulation of collagen's conformational behaviour is not due to changes in flexibility, but rather arises from the induction of curvature (either intrinsic or induced by interactions with the mica surface). By modifying standard polymer theory to include innate curvature, we show that collagen behaves as an equilibrated curved worm-like chain (cWLC) in two dimensions. Analysis within the cWLC model shows that collagen's curvature depends strongly on pH and salt, while its persistence length does not. Thus, we find that triple-helical collagen is well described as semiflexible, irrespective of source, type, pH and salt environment. These results demonstrate that collagen is more flexible than its conventional description as a rigid rod, which may have implications for its cellular processing and secretion.

## INTRODUCTION

Collagen is the predominant structural protein in vertebrates, where it represents more than one quarter of the total protein in our bodies (1, 2). It is widely used as a biomaterial, and plays a vital physiological role in extracellular matrix and connective tissue mechanics. Collagen can assemble into many different higher-order forms, which fulfil distinct structural and mechanical roles.

Over 28 different types of human collagen have been identified (1, 2), of which the most prevalent are fibrillar collagens such as types I, II and III. These collagens assemble to create highly ordered fibrils, which in turn are the building blocks for the extracellular matrix and for fibres, which act as load- and tension-bearing structures in connective tissues. Not surprisingly, changes in collagen's composition are associated with a wide variety of diseases, including osteogenesis imperfecta (3) and Ehlers-Danlos syndrome (4). In addition to genetic mutations in collagen, physiological dysfunction can arise from alterations in posttranslational modifications, and aging via nonenzymatic glycation and crosslink formation (5). Such chemical changes at the protein level correlate with altered tissue structure and mechanics, and pathologically affect human health. Because of the hierarchical nature of collagen structure, identifying the mechanisms by which chemical changes modify tissue mechanics requires understanding how they impact mechanics at the molecular level.

Collagen's molecular structure is a right-handed triple helix (Figure 1A), approximately 300 nm long and 1-2 nm in diameter (2). This triple helix comprises three left-handed polyproline-II-like helices (α-chains) with a characteristic (Gly-X-Y)$_n$ repeat amino acid sequence. Although frequently proline or hydroxyproline, the X and Y amino acids are variable, and provide the sequence diversity that defines individual collagen types. Different collagen types are also distinguished as homotrimeric (three identical α-chains) or heterotrimeric. Although the sequences of the different types of collagen are well established, their mechanical properties are not.

One property commonly used to describe the mechanical properties of chain-like biomolecules is the persistence length – the length over which the orientation of the chain remains correlated in the presence of thermal noise. While the persistence lengths of other biological polymers such as DNA are well established and robust to experimental technique, this is not the case for collagen (Table 1). Early solution-based studies found that collagen's triple helix should be considered semi-rigid, exhibiting a persistence length of $p$ ≈ 130-180 nm (6-8). These values likely led to the conventional description of collagen as a possessing a stiff, rod-like structure (9, 10). However, most recent single-molecule experimental and simulation studies contradict this description, finding persistence lengths for collagen as short as $p$ ≈ 10 nm (11-17). These values imply that collagen is highly flexible, adopting compact, coiled configurations in solution. Other experimental and simulation approaches have determined persistence lengths throughout this range, with another cluster of values around $p$ ≈ 40-60 nm (14, 18-20). It is remarkable that this fundamental parameter of persistence length is so poorly established for a protein of such mechanical importance and ubiquity.

**Table 1** – Literature estimates of persistence length for fibrillar, molecular collagens.

| Persistence length, $p$ (nm) | Collagen type and source | Solution conditions | Method | Reference |
|---|---|---|---|---|
| 10 | 57-residue segment of homotrimeric mouse type I | water, neutral pH | Molecular dynamics | (16) |
| 12 | Bovine dermis type I | water, pH~5 | AFM imaging | (13) |
| 13 | 45-residue segment of rat type I | 10 mM NaCl, neutral pH | Molecular dynamics | (17) |
| 11-15 | Cell-derived human type I procollagen; recombinant human type II procollagen | >10 mM buffer, neutral pH | Optical tweezers stretching | (11, 12) |
| 16 | 30-residue homotrimeric peptide | water, neutral pH | Steered molecular dynamics | (15) |
| 22 | 57-residue segment of heterotrimeric mouse type I | water, neutral pH | Molecular dynamics | (16) |
| 15-65 | Recombinant human type II procollagen | >10 mM buffer, neutral pH | Optical tweezers stretching | (14) |
| 40 | Fetal bovine type III | 50 mM acetic acid + equal volume glycerol | Electron microscopy | (18) |
| 51 | 60-nm segment of human type I collagen | water, neutral pH | Coarse-grained molecular dynamics | (20) |
| 57 | Calf dermis type I | 50 mM acetic acid + equal volume glycerol | Electron microscopy | (18) |
| 130 | Rat skin type I | >10 mM buffer, pH ~5 | Viscometry | (6) |
| 135-165 | Bovine dermis type I | >10 mM buffer, neutral pH | AFM imaging | (13) |
| 161 | Rat skin type I | 0.3 M acetate+ 6 mM NaCl, pH 4 | Rheology | (7, 8) |
| 160-165 | Bovine dermis type I | 1 mM HCl | Dynamic light scattering | (19) |
| 167 | Rat skin type I | >10 mM buffer, neutral pH | Rheology | (8) |

Examination of Table 1 suggests that there is diversity not only in the approaches used to study collagen's flexibility but also in the types of collagen, their sources, and solution conditions. A comparison of different tissue-derived collagen types was performed in reference (18), which determined the type I heterotrimer to be slightly more rigid than the type III homotrimer. A similar difference between homo- and heterotrimeric collagen sequences was found in recent MD simulations (16). Collagen's remodelling by MMP1, a collagenase that unwinds the triple helix, has been suggested to depend on the source of collagen (e.g. recombinant, from cell culture or from tissue) (21), implying that posttranslational enzymatic and age-related nonenzymatic modifications may alter its mechanics at the molecular level. A recent study found that collagen's flexibility is influenced by its solution environment, with collagen appearing flexible at low salt in acidic conditions and more rigid in high-ionic-strength, neutral pH buffers (13). This result runs counter to the expectations of the behavior of polyelectrolytes like DNA, which are expected to become more flexible as ionic strength is increased (22). Understanding collagen's response in different solution conditions is relevant not only to its mechanical role in distinct tissue environments, but also to understanding its intracellular compactness as it traverses a pH gradient during cellular processing and secretion (23, 24).

In this work, we use atomic force microscopy imaging and characterization to investigate how composition, source and chemical environment affect the flexibility of collagen's triple helix. By comparing different fibrillar collagens (types I, II and III; tissue-derived and recombinantly expressed; homo- and heterotrimeric triple helical structures), we find that molecular composition does not significantly impact collagen's overall flexibility. In contrast, we find that both ionic strength and pH independently impact the apparent flexibility of collagen, providing estimates of persistence length that span the range from flexible to semi-rigid, depending on solution environment. Careful consideration of polymer chain statistics shows, however, that treating collagen as a standard worm-like chain does not describe its properties in most of these chemical environments. Instead, we show that a curved worm-like chain model that includes inherent molecular curvature is a far more appropriate descriptor of the observed conformations of collagen on mica. With this model, collagen's bending flexibility depends much less on solution conditions; instead, salt concentration and pH modulate its global curvature. We find that curvature does not depend strongly on the type or source of collagen, nor does the persistence length. To the best of our knowledge, these results constitute the first experimental analysis of curved worm-like chains, and provide a new explanation for the conflicting reports on the flexibility of triple-helical collagen.

**MATERIALS AND METHODS**

**Collagen Sources**

Recombinant human type I (RhC1-003) and recombinant human type III (RhC3-012) collagen were expressed in yeast and obtained from FibroGen (generous gifts of Alexander Dunn, Stanford University), rat tail tendon-derived type I collagen (3440-100-01) was purchased from Cultrex, and human cartilage-derived type II collagen (CC052) was purchased from EMD Millipore. All stocks are between 2 and 5 mg/ml in 20 mM acetic acid.

**Sample Preparation**

The desired solution conditions were obtained by solution exchange using Millipore Amicon Ultra-0.5 spin filters (NMWL 50 kDa, UFC505096), then dilution to approximately 1 µg/ml collagen prior to deposition. 50 µl was deposited onto freshly cleaved mica (Highest Grade V1 AFM Mica Discs, 10 mm, Ted Pella) for 20 seconds. After deposition, samples were rinsed five times with 1 ml ultrapure water to remove unbound proteins, and the mica was dried under a flow of filtered compressed air. It is important to note that all collagen molecules were imaged in these dry conditions. Thus, solution conditions quoted refer to the condition under which collagen was deposited onto mica, at room temperature.

**Atomic Force Microscopy Imaging**

Images of collagen adsorbed to mica were collected with an Asylum Research MFP-3D atomic force microscope using AC tapping mode in air. AFM tips with a 325 kHz resonance frequency and 40 N/m force constant (MikroMasch, HQ:NSC15/AL BS) were used for image collection, and were changed as necessary to preserve image quality.

**Chain Tracing**

The SmarTrace algorithm used to trace collagen chains from AFM images was developed in MATLAB (25) and uses a graphical user interface adapted from (26). It is available from the authors upon request. A detailed description of the SmarTrace workflow and validation can be found in the Supporting Information. Briefly, the algorithm uses cross-correlation analysis of a template cross-section with the imaged chain in order to identify its center line and width. Continuity constraints on the local directionality and width enable the tracing of chains in noisy environments.

**Data Analysis**

Traced chains were sampled using bootstrapping to extract statistics of the chain for flexibility determination. Following the method of Faas *et al.* (27), each chain was randomly divided into non-overlapping segments of lengths drawn from a set of input values (here, $s = 10$ nm, 20 nm, 30 nm, ... , 200 nm). The maximum segment length does not need to be the same as the chain contour length, so it allows the use of partially traced chains. This is particularly useful when ends of a molecule are not clear or chains intersect, as it allows subsections of the chain to be included in the analysis. $\langle R^2(s) \rangle$ and $\langle \cos \theta(s) \rangle$ were determined from all segments of each length $s$.

Resultant $\langle R^2(s) \rangle$ and $\langle \cos \theta(s) \rangle$ were fit to equations derived from the inextensible worm-like chain (WLC) model with (equations 3-4) or without (equations 1-2) intrinsic curvature. Derivations for equations 3 and 4 are presented in the Supporting Information.

The kurtosis of each angular distribution was calculated at each segment length $s$ as $n[\sum_i (\theta_i - \langle \theta \rangle)^4] / [\sum_i (\theta_i - \langle \theta \rangle)^2]^2$, for a distribution across $n$ angle bins $\theta_i$. The standard error of the kurtosis was determined using Eq. S19.

Values for $p$ and $\kappa_o$ presented in Table 2 are an average of the results from independent fits to $\langle R^2(s) \rangle$ and to $\langle \cos \theta(s) \rangle$. Reported errors, $\Delta$, represent the propagated error of the 95% confidence intervals of the respective fit parameters or half of the difference between the $\langle R^2(s) \rangle$ and $\langle \cos \theta(s) \rangle$ fit parameters, whichever is larger. A full list of fitting parameters from all models and samples is provided in Table S1. The total length of collagen chains traced in each sample condition is also included in Table S1.

**Validation Tests**

The performance of the SmarTrace algorithm and data analysis code was evaluated by tests on simulated polymers, and on images of DNA. Details are provided in the supporting information.

RESULTS AND DISCUSSION

Atomic force microscopy (AFM) is a well-established tool for imaging the conformations of flexible biopolymers (28). We have used the technique to image collagens of different types and sources, deposited from a range of solution conditions, and have characterized how these parameters influence the flexibility of collagen's triple helix.

**Collagen type and source**

To investigate how collagen's flexibility depends on its source, we compared four different samples of collagen. These represent three types of fibrillar collagens (types I, II and III), of either rat (type I) or human (all three) genetic origin. Although all of these samples are capable of self-assembly into highly ordered fibrils, they were imaged under sufficiently dilute conditions such that assembly did not occur. Type I collagen is a heterotrimer ($\alpha_1(I)_2\alpha_2(I)_1$), while types II and III are each homotrimeric ($\alpha_1(II/III)_3$). Additionally, to explore how collagen's source influences its flexibility, our samples encompass both tissue-derived (rat type I and human type II) and cell-derived (human types I and III) sources. The latter were recombinantly expressed in yeast, and possess only prolyl hydroxylation as a posttranslational modification (PTM) (29). This contrasts with mammalian-produced collagens which have extensive PTMs including hydroxylation of prolines and lysines, as well as O-glycosylation of hydroxylysines (14, 23).

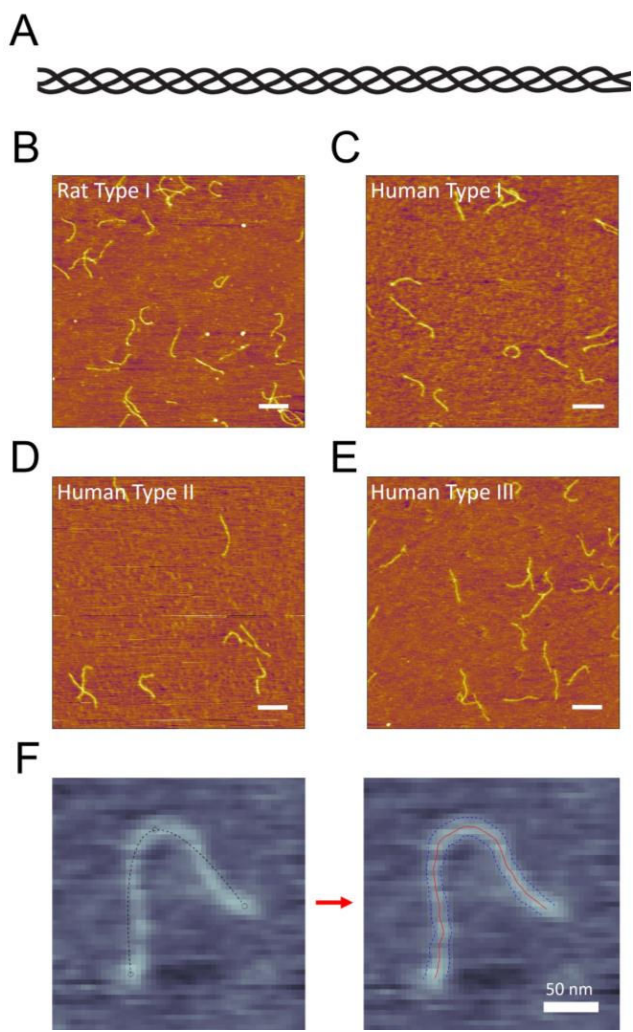

**Figure 1**. Imaging individual collagen proteins. (A) Schematic of triple-helical collagen. (B-E) Representative AFM images of each of the four collagen samples investigated in this work. For each sample, collagen was deposited from a solution of 100 mM KCl + 1 mM HCl. Scale bars = 250 nm. (B) Rat tail-derived type I collagen. (C) Recombinant human type I collagen. (D) Cartilage-derived human type II collagen. (E) Recombinant human type III collagen. (F) SmarTrace starts with user input points along a collagen chain (left image). Initial splines connecting the three input points (dashed lines) do not follow the chain. The SmarTrace algorithm identifies both the centreline of the chain (red) and its width (blue – used for chain tracing but no further analysis), as shown in the right image.

Representative AFM images of each of the four collagen samples are shown in Figure 1, B-E. Each image was obtained by depositing collagen from a room-temperature solution of 100 mM KCl + 1 mM HCl onto freshly cleaved mica, then rinsing with water and drying prior to imaging. For all four collagen samples, the chains appear semiflexible, with contour lengths of approximately 300 nm, as expected. There are no obvious qualitative differences in flexibility among these samples.

To quantify the flexibility of each collagen type, the imaged chains were traced to provide backbone contours for conformational analysis. Existing chain-tracing algorithms proved problematic for many of our images, in some cases due to the background levels of noise (27), or in others from the sensitivity of the algorithm to the locations of user-input starting points (26). Hence, we developed a new chain-tracing program, dubbed SmarTrace. The algorithm incorporates pattern matching to identify the best centerline of a chain, and refines an initial guess at this centerline by using direction and width continuity constraints. Required user input is minimal: backbone contours are identified from only a few clicks near the chain (see Figure 1F). Details of SmarTrace's methodology and workflow are provided in the Supporting Information (SI).

From traced chains, we implemented tools of polymer physics to analyse the statistical properties of chain conformations and determine persistence lengths (30). To perform this analysis, chains were segmented randomly into non-overlapping pieces of different contour lengths (e.g. $s = 10, 20, \ldots, 200$ nm). This approach allows partially traced chains to be included in the analysis (27). For each segment, we calculated its squared end-to-end distance $R^2(s)$ and the change in orientation between its starting and ending tangent vectors $\hat{t}(s) \cdot \hat{t}(0) = \cos\theta(s)$; these quantities were then averaged over all segments of length *s* within the population. Length-dependent trends in mean squared end-to-end distance, $\langle R^2(s) \rangle$, and tangent vector correlation, $\langle \cos\theta(s) \rangle$, were compared with the predictions of the worm-like chain (WLC) model for polymers equilibrated in two dimensions (2D):

$$\langle R^2(s) \rangle = 4sp \left[ 1 - \frac{2p}{s}\left(1 - e^{-\frac{s}{2p}}\right) \right] \quad (1)$$

$$\langle \cos\theta(s) \rangle = e^{-\frac{s}{2p}}. \quad (2)$$

The validity of the SmarTrace chain-tracing algorithm and analysis approaches was established by testing their performance on AFM images of DNA and on simulated chains (see SI and Figures S1-S3).

We first investigated whether collagen molecules are equilibrated on the mica and are well described by the worm-like chain model. For type I collagen from rat deposited from 100 mM KCl + 1 mM HCl, both mean-squared end-to-end distance $\langle R^2(s) \rangle$ and tangent vector correlation $\langle \cos\theta(s) \rangle$ are well described by the WLC model (Figure 2A-B). The agreement between persistence lengths from the fits to equations (1) and (2), $p = 116 \pm 3$ nm and $p = 107 \pm 6$ nm respectively, further suggest that collagen deposited from this solution condition behaves as an equilibrated 2D WLC. This assumption is corroborated by Gaussian distributions of bending angles (Figure 2C). The kurtosis of the bending angle distribution is approximately 3 (the value characteristic of a normal distribution) at all contour lengths (Figure 2D). These measures all indicate that the conformations of collagen on mica represent a sample equilibrated in two dimensions, and that rat type I collagen has a persistence length of approximately 110 nm.

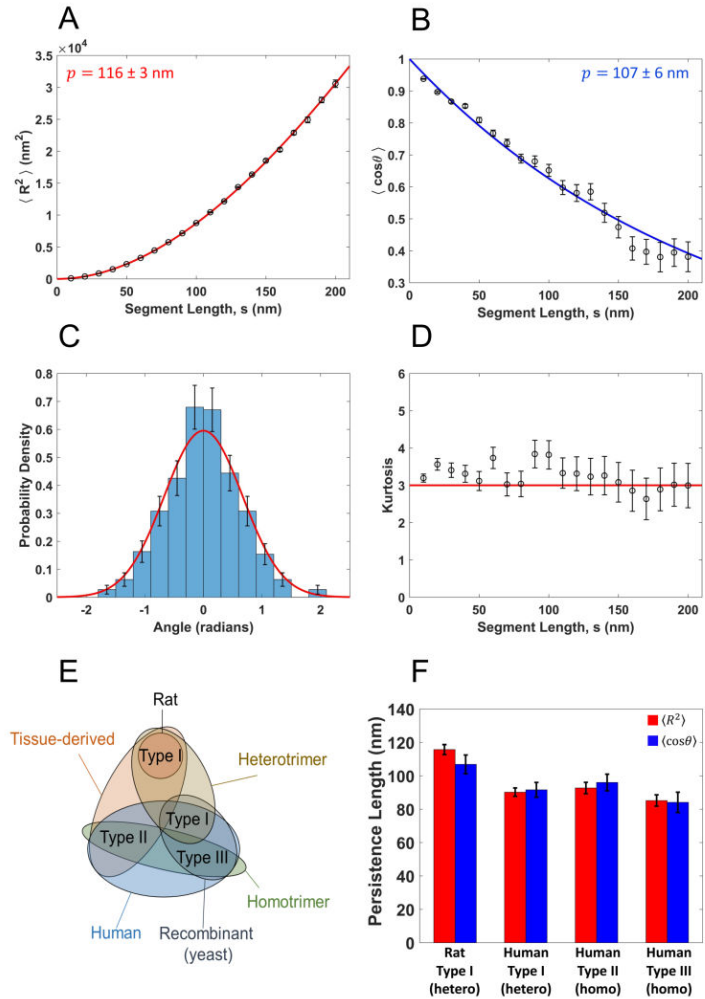

**Figure 2**. Determination of collagen's persistence length using the standard worm-like chain (WLC) model. Parts (A-D) present analysis for rat type I collagen deposited from a solution of 100 mM KCl + 1 mM HCl. (A) Mean squared end-to-end distance as a function of segment length, $\langle R^2(s) \rangle$. Points represent mean values determined at 10 nm segment-length intervals, with error bars representing standard errors of the mean. The red line is a fit with equation 1, yielding a persistence length of $p = 116 \pm 3$ nm (error represents 95% confidence interval of the fit). (B) Tangent vector correlation as a function of segment length, $\langle \cos\theta(s) \rangle$. Points and errors are similarly represented as in (A). The blue line is a fit with equation 2, yielding $p = 107 \pm 6$ nm. Fits to the WLC model in (A) and (B) capture the trends in the data and yield comparable persistence lengths. (C) Angular histogram for 50 nm segment lengths (blue bars). Error bars represent $\sqrt{N}$ counting error. This distribution agrees well with that expected for a worm-like chain of persistence length $p = 112$ nm (the average of the results from (A) and (B)), which is a normal distribution with a mean of zero and a variance of $s/p$ (red line). (D) Kurtosis of the angular distributions extracted from the traced collagens at different segment lengths. Error bars represent the standard error in the kurtosis (Eq. S19). The kurtosis of the angular distributions is close to 3 (red line) for all segment lengths, indicating that the collagens behave as equilibrated Gaussian chains on the mica surface. (E) Venn diagram illustrating the similarities and differences between the different collagen samples tested, including trimeric identity and source. (F) Persistence length for each collagen sample deposited from a solution of 100 mM KCl + 1 mM HCl, obtained using $\langle R^2(s) \rangle$ and $\langle \cos\theta(s) \rangle$ analyses. The similarity among these samples suggests that collagen type and source have little impact on the mechanical properties of collagen at the molecular level.

The persistence lengths were similarly determined for the other collagen samples (Figure 2E). The data and fits are shown in Figure S4, while persistence lengths are presented in Figure 2F. Table 2 provides a summary of the fit parameters, representing an average of the $\langle R^2(s)\rangle$- and $\langle \cos\theta\,(s)\rangle$-derived results; the complete summary of parameters from each fit is given in Table S1, along with reduced $\chi^2$ values used to assess goodness of fit. All four collagen samples exhibit similar persistence lengths, within the range of approximately 80-120 nm. We find no significant differences in flexibility between tissue-derived (rat I, human II) and yeast-expressed recombinant (human I, III) collagens. We observe slightly greater flexibility for homotrimeric (II, III) collagen versus heterotrimeric (I) collagen, similar to previous findings (18, 31). However, these differences are small in the context of the wide range of values in the literature (Table 1). Thus, differences in type, source and extent of posttranslational modifications do not appear to cause substantial differences in molecular flexibility.

**Influences of salt and pH**

Given a previous report of significantly different collagen flexibility in buffers with high salt concentration and neutral pH, compared with low ionic-strength, acidic solutions (13), we next investigated how pH and salt concentration independently impact the flexibility of collagen. Rat type I collagen was deposited onto mica from a range of different solution conditions in a controlled ionic environment. In one set of experiments, solutions contained only varying concentrations of potassium chloride, from 0 mM (water) to 100 mM KCl. In a second set of experiments, solutions again contained varying concentrations of KCl, but were acidified to pH ≈ 3 by 1 mM HCl.

Images of collagen at different KCl concentrations and pH are shown in Figure 3A. It is immediately apparent that salt affects the conformations of collagen: the protein transitions from more compact structures at low ionic strength to much more extended conformations at high ionic strength. A similar trend is seen in the acidic conditions. Conformations were analysed to obtain mean-squared end-to-end distances and tangent correlations as a function of contour length, and were fit to equations (1) and (2), respectively. This analysis finds persistence length to increase significantly with ionic strength, at both neutral and acidic pH (Figure 3B and Table 2). Our results corroborate and elaborate on the previous findings of increased flexibility at low pH and salt compared with neutral pH and high salt (13).

The dependence of persistence length on salt concentration is striking. In neutral solutions, the determined persistence length exhibits a four-fold increase from its value in water as the KCl concentration is increased to 100 mM. In acidic solutions, persistence lengths are shorter but still increase, by about three-fold as the concentration of KCl increases from 1 mM to 100 mM. Intriguingly, the range of persistence lengths exhibited here spans almost the entire range of values reported, and unreconciled, in the literature (Table 1).

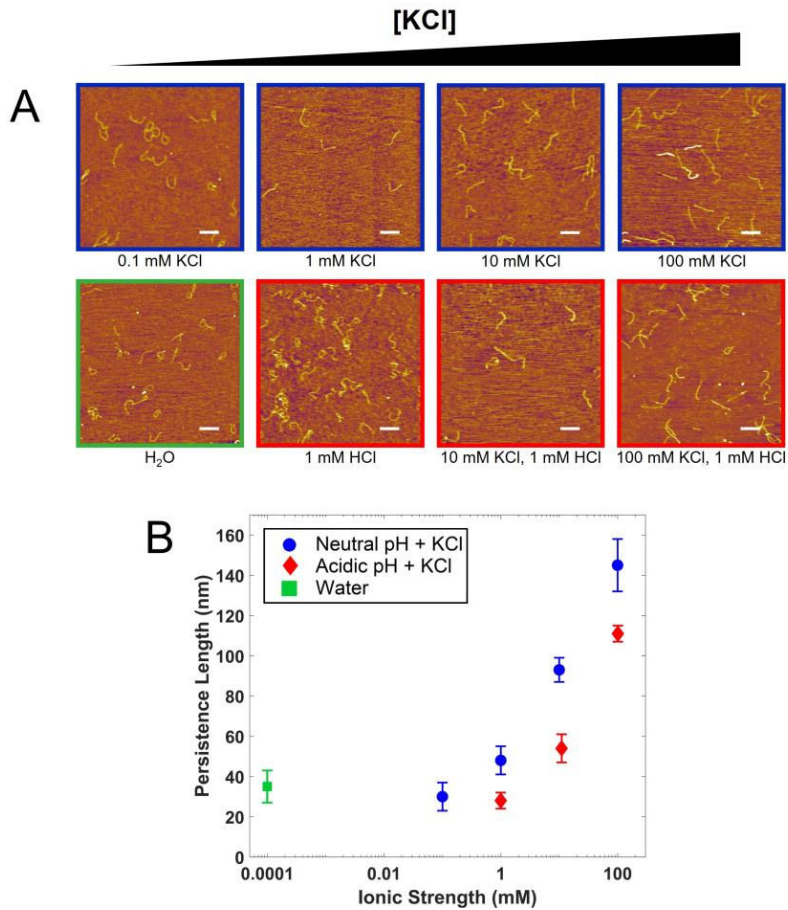

**Figure 3**. Effects of ionic strength on persistence length. (A) AFM images of rat type I collagen deposited from solutions covering a range of KCl concentration and pH. All scale bars represent 250 nm. As salt concentration increases, the collagen chains appear to straighten. (B) Persistence lengths in these conditions. The reported values are the average of $p$ extracted from fits to $\langle R^2(s) \rangle$ and to $\langle \cos\theta(s) \rangle$, and errors are the larger of the propagated error in this mean or half the separation between the two values of $p$. For the purposes of graphical representation, water was assumed to have an ionic strength of $10^{-7}$ M. Increasing the ionic strength causes a large increase in the apparent persistence length, an effect that is reduced slightly in the acidic condition for the same ionic strengths.

The trend of increasing stiffness with increasing ionic strength is unexpected, as it is opposite to predictions for polyelectrolytes and to observations for biopolymers such as DNA (22). However, examination of the data and fits in some of the solution conditions, particularly at lower salt concentrations, reveals that the experimental data are not well described by the standard worm-like chain model (e.g. Figure 4; see also Figures S5, S6). For $\langle R^2(s) \rangle$, the agreement between model and data is reasonable at shorter segment lengths (e.g. $s$ < 100 nm), but longer segments exhibit shorter-than-expected end-to-end distances. The disagreement with the standard WLC model is even more apparent when considering the $\langle \cos\theta(s) \rangle$ behaviour. There appears to be an oscillatory modulation of the tangent vector correlation, whereby chains maintain directionality at small segment lengths, reorient substantially at intermediate lengths, and in many cases, exhibit anticorrelated tangent vectors at longer contour lengths. These are not features of a well equilibrated WLC. Instead, the oscillatory character of the tangent vector decorrelation suggests that curvature may play a role in the observed conformations of collagen.

**Curved worm-like chain**

To describe an intrinsically curved worm-like chain (cWLC), we assume that the lowest-energy conformation of the chain is curved (bent), rather than straight. The extent of angular fluctuations about this curved state is determined by the angular bending potential and the thermal energy of the system, which relate to persistence length as for the standard worm-like chain. From this cWLC model, equations for the mean squared end-to-end distance and mean tangent vector correlation can be derived (see Supporting Information):

$$\langle R^2(s) \rangle_c = \frac{4sp}{(1+4\kappa_0^2 p^2)^2} \left\{ 1 - \frac{2p}{s}(1 - 4\kappa_0^2 p^2)\left[1 - \cos(\kappa_0 s)\, e^{-\frac{s}{2p}}\right] + \frac{4\kappa_0 p^2}{s}\left[\kappa_0 s - 2\sin(\kappa_0 s)\, e^{-\frac{s}{2p}}\right] \right\}, \quad (3)$$

$$\langle \cos\theta(s) \rangle_c = \cos(\kappa_0 s)\, e^{-\frac{s}{2p}}. \quad (4)$$

Here $\kappa_o$ represents the inherent curvature of the chain, which is the inverse of its inherent radius of curvature: $\kappa_o = R_o^{-1}$. Equations (3) and (4) describe the expected behavior of an intrinsically curved, inextensible worm-like chain equilibrated in two dimensions, and are two-dimensional versions of 3D results derived previously (32). Of note, the cWLC model is distinct from approaches that assume a random local curvature (27): in the cWLC model the lowest-energy conformation of the chain is globally curved, *i.e.*, bends always in the same direction.

Figures 4A and 4B show fits of the curved WLC model to the $\langle R^2(s) \rangle$ and $\langle \cos\theta(s) \rangle$ data from rat type I collagen deposited from 1 mM HCl. The cWLC captures trends in both $\langle R^2(s) \rangle$ and $\langle \cos\theta(s) \rangle$ significantly better than the standard model ($\chi^2_{red}$ for $\langle R^2(s) \rangle$: 18 (cWLC) vs 423 (WLC); for $\langle \cos\theta(s) \rangle$: 4.4 (cWLC) vs 68 (WLC)). Notably, the cWLC model captures the longer-length anticorrelation of the tangent vectors ($\langle \cos\theta \rangle < 0$), which under the standard WLC model is unphysical. Additionally, the cWLC fits provide a distinct interpretation of the compact structures observed at low ionic strength (Figure 3): rather than being more flexible at lower salt concentration, collagen is instead more curved.

Salt-dependent trends in persistence length and curvature obtained from the curved WLC model are shown in Figures 4C and 4D. In contrast to the monotonic increase in persistence length with increasing KCl concentrations found when using the standard WLC model (Figure 3), the results from the cWLC fits indicate that persistence length varies only modestly over the range of concentrations used here (Figure 4C). Instead, the curvature depends strongly on ionic strength, decreasing from $\kappa_0 = 0.02$ nm$^{-1}$ to 0 (*i.e.*, straightening) as the concentration of KCl is increased (Figure 4D). This effect is shown schematically in Figure 4E.

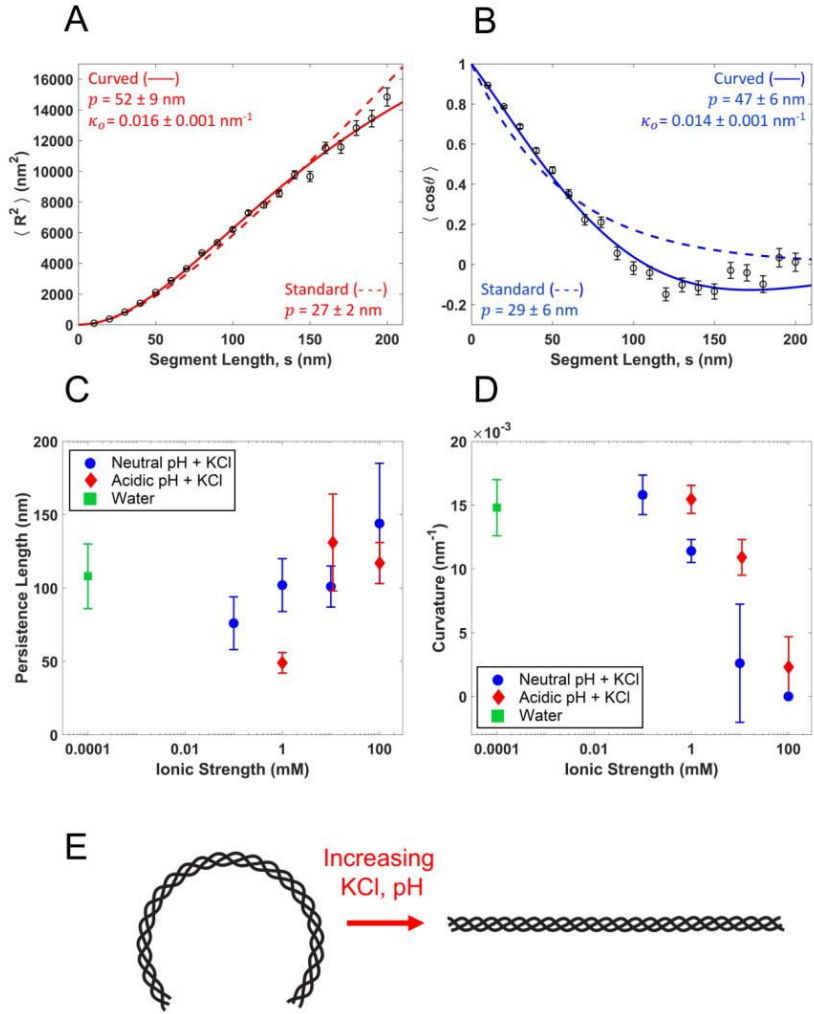

**Figure 4**. At low ionic strength, collagen is better described as a curved worm-like chain. (A) Fits of the mean squared end-to-end distance using both the standard (Eq. 1, dashed red line) and curved (Eq. 3, solid red line) WLC models, to data extracted from rat tail type I collagen deposited from 1 mM HCl. (B) Corresponding fits of the standard (Eq. 2, dashed blue line) and curved (Eq. 4, solid blue line) WLC tangent vector correlation functions. The data are more accurately described by the curved model equations, as shown statistically by a significant improvement in reduced $\chi^2$ values (see text). The standard WLC underestimates the persistence length of the collagens in this condition, presumably interpreting the induced curvature as additional fluctuation. (C) Persistence lengths for the different co-solute conditions presented in Fig. 3. Rather than the trend with ionic strength seen with the standard WLC fits, there is a less significant, and less obviously monotonic, variation in persistence length as a function of ionic strength. Instead, the observed systematic conformational changes are attributed to a variation in innate curvature, which drops as the ionic strength of the solution is increased (D). For the purposes of graphical representation, water was assumed to have an ionic strength of $10^{-7}$ M. Error bars on $\langle R^2(s) \rangle$ and $\langle \cos\theta(s) \rangle$ values are as described for Figure 2, and error bars on $p$ and $\kappa_o$ are obtained as described for Figure 3. (E) Schematic illustrating the transition of collagen from a molecule with a curved backbone at low ionic strength and pH (left) to one with a straight backbone at higher ionic strength and neutral pH (right). The persistence length characterizes fluctuations about this lowest-energy conformation.

**Origins of collagen curvature**

To determine whether the induction of this curvature is ion-specific, images of rat type I collagen deposited from 20 mM acetic acid were collected and analyzed with the standard and curved WLC models (Figure 5A-D). Conformations of collagen deposited from acetic acid are better described by the cWLC model, consistent with observations at low ionic strengths of KCl. Again, naïve application of the standard WLC model results in persistence lengths smaller than those obtained from cWLC fits. To test for equilibration of the samples, we considered the distribution of bend angles. Because of chain curvature, these distributions should not be represented by a single Gaussian centered at $\theta(s) = 0$. Rather, the distributions are expected to be described by two Gaussians, centered symmetrically about zero at an angle $\theta(s) = \pm\kappa_0 s$ and with standard deviation $\sigma(s) = \sqrt{s/p}$ (equation (S20)). The experimental angular distribution is indeed well described by this bimodal function (Figure 5D). The agreement between predicted and measured angular distributions is strong evidence of collagen equilibration on the mica surface in this low-salt condition. The fit parameters in acetic acid ($p = 50$ nm, $\kappa_o = 0.018$ nm$^{-1}$) are similar to those from the 1 mM HCl solution (Table 2), which has comparable ionic strength and pH. Thus, at low ionic strength, there does not seem to be a strong effect of chloride versus acetate anions on collagen's mechanical properties.

It is possible that the observed curvature is an effect specific to heterotrimeric collagen. Since type I collagen has one α-chain which is different than the other two, a larger or smaller propensity for surface interactions with this chain could produce a surface-induced curvature to minimize its energy (33). To address this possibility, types I, II and III human collagens were also imaged in 20 mM acetic acid and analyzed with the curved WLC model (Figure S7). The results of this analysis are shown in Figures 5E and 5F. Although the different types vary somewhat in their persistence lengths, their curvatures are nearly identical under these solution conditions, regardless of their trimeric identity. Thus, surface interactions with a unique chain in heterotrimeric collagen are not responsible for the observed curvature.

The curvature observed at low ionic strengths may nonetheless result from molecular interactions with the surface. Intrinsically straight, chiral semiflexible chains may adopt curved conformations at an interface (33). The extent to which the chain curves at an interface will be determined by a competition between the free energy of surface-molecular interactions and the torsional twist energy of the chain. Specifically, curvature in this model requires a chiral organization of sites on the chain that interact differentially at the interface compared with the rest of the chain sites. In the current context, these could be the side-chains of the three individual α-chains of collagen, which are arranged in a right-handed helical structure about the central axis of the collagen molecule. If these sites interact strongly with the mica surface, the free energy gained by adhering to the surface could outweigh the energetic cost of altering the twist (local helical pitch) of the triple helix. In this scenario, the observed trends in curvature with salt concentration (Figure 4F) imply a salt dependence of collagen's torsional stiffness and/or its interactions with the surface. Salt-dependent interactions with mica are well established for other biopolymers (e.g. (30)) and could be expected to contribute here (34). In particular, K$^+$ affects the interactions of self-assembling collagen with a mica surface (34-36). However, KCl also affects collagen stability in solution: the thermal stability decreases at lower concentrations of KCl (37). If interactions with the surface contribute to the observed low-ionic-strength curvature, destabilization of the triple helix at low salt concentration – seen also by decreased circular dichroism of the triple helix (37) – is consistent with a reduced torsional stiffness of collagen.

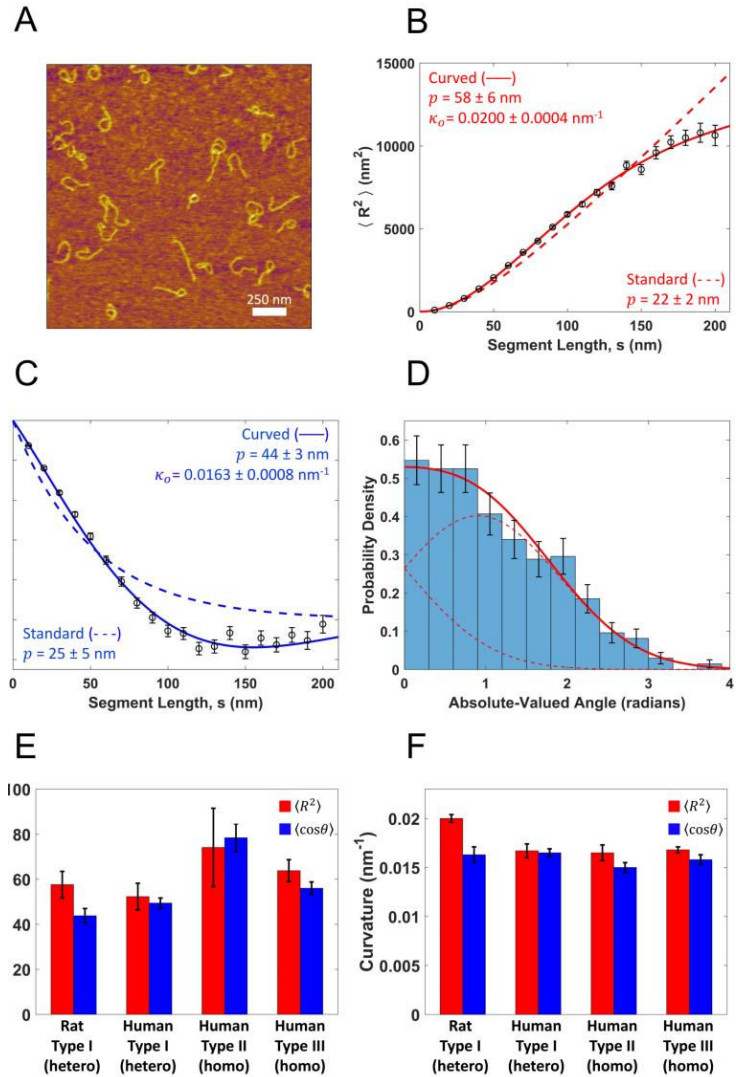

**Figure 5**. Behavior of different collagen types in acetic acid. (A) Representative image of rat tail type I collagen deposited from 20 mM acetic acid. (B) Standard (Eq. 1, dashed red line) and curved (Eq. 3, solid red line) WLC model fit to the mean squared end-to-end distance, $\langle R^2(s) \rangle$. (C) Standard (Eq. 2, dashed blue line) and curved (Eq. 4, solid blue line) WLC model fit to the tangent vector correlation, $\langle \cos\theta(s) \rangle$. Data in both (A) and (B) are described more accurately by the curved WLC model, as seen also by the reduced $\chi^2$ values (Table S1). (D) Absolute-valued angular distribution of the same collagen molecules at a segment length of $s = 50$ nm. The red line represents the expected distribution (Eq. S3) for a curved WLC using a persistence length $p$ = 51 nm and intrinsic curvature $\kappa_o$ = 0.0182 nm$^{-1}$, the average values of each parameter extracted from the fits shown in (B) and (C). The dashed red lines are the two Gaussian components which make up this distribution, as discussed in the SI. (E) and (F) Persistence lengths and curvatures, respectively, of different collagen types deposited from 20 mM acetic acid. As with the different collagen types deposited from 100 mM KCl + 1 mM HCl (Figure 2F), there is only modest variation amongst the different samples, demonstrating that, also at lower ionic strength and different co-solute conditions, only minor mechanical differences exist between different collagen types and sources.

In addition to the ionic-strength dependence of curvature, our images and conformational analysis also demonstrate a dependence on pH: collagen appears more curved in more acidic conditions (Figure 4F). At a pH of 3, collagen carries a net positive charge. At neutral pH, collagen instead has a roughly equal balance of positively and negatively charged residues. The increased curvature at acidic pH could result from stronger electrostatic affinity to the negatively charged mica surface (34). Alternatively, electrostatic repulsion between α-chains could destabilize the triple helix and reduce its torsional stiffness, particularly at low ionic strength when shielding from co-solute anions is weak (at 1 mM ionic strength the Debye length ≈ 10 nm). At neutral pH, the potential for salt bridge formation could strengthen attractions between α-chains (37, 38), thereby increasing the torsional stiffness of the triple helix.

It is also possible that collagen itself is intrinsically bent, even in solution (39). Certainly within the fibrillar superstructure, collagen molecules display conformations that are locally bent (40). Studying how salt influences conformations in solution would be key to discriminating which interactions are surface-specific in our study. A change in torsional stiffness and/or alteration of preferred twist angle at lower ionic strength would be expected to alter the average shape of collagen in solution. Light-scattering or other solution-based techniques could be used to study how collagen's compactness in solution depends on salt concentration, and thereby determine how the triple helix itself responds to changes in ionic strength.

**Comparison with other estimates of collagen flexibility**

Taken together, our results indicate that collagen should be regarded as a semiflexible polymer at room temperature, with a persistence length in the range of 80-100 nm. This is more flexible than solution-based estimates, yet more rigid than found in almost all previous single-molecule approaches (Table 1). Inherent biases and assumptions built into data interpretation within each methodology may contribute to the range of results reported in the literature. Here, we describe possible reasons for the discrepancies.

With AFM imaging, one must play particular heed to molecule-surface interactions and to the possibility of kinetically trapping polymers in non-equilibrium configurations on the surface. In our measurements, the observed conformations appear well equilibrated in two dimensions (Figure 2C,D and Figure 5D). Furthermore, while electrostatic interactions with mica may play a role in the apparent curvature of collagen on the surface, we have found that the persistence length is not strongly affected.

In single-molecule stretching experiments, one source of bias is the underestimation of persistence length when the contour length $L < 20p$ (41), as is the case for collagen. Any structural changes in the triple helix that occur during stretching may also modify the bending energy: such changes have been inferred from enzymatic cleavage assays of collagen under force (21, 42-44), and may contribute to softening of collagen's force-extension response (14).

Many of the simulations of collagen have used steered molecular dynamics, which may not investigate its equilibrium flexibility. Additionally, the results of molecular dynamics simulations depend on the force fields used (45), which have been developed predominantly for globular proteins rather than the unique fold of collagen.

Interpretations of light-scattering data require a model (such as an ellipsoidal scatterer (19)), and it is possible to describe decay curves equally well assuming both compact and more extended molecular geometries (46). Furthermore, the effective size of the protein, as deduced from solution-based studies such as light-scattering and rheology, is often affected by intermolecular interactions; thus, concentration-dependent studies are essential in order to determine the spatial extent of a given molecule in isolation (47). Given collagen's propensity to self-associate (48, 49), intermolecular interactions may contribute to larger estimates of collagen's size in solution. The rheological studies of reference (8) found that pH does not have a significant effect on collagen's flexibility; however, to reach this conclusion, the authors had to assume that type I collagen exhibited a non-uniform flexibility at neutral pH, with ends considerably more flexible than the central region of the triple helix.

The two studies most similar to ours, which used AFM imaging (13) and electron microscopy (EM) (18) to analyse collagen's flexibility, based their estimates of persistence length on the distance between the endpoints of the collagen chain. As we have shown (Figures 4 and 5), naïve application of equation (1) can lead to significant underestimation of persistence length for curved chains. Consistently, our estimates of collagen's persistence length are greater than theirs obtained at low ionic strength. At high ionic strength and neutral pH, our results and those of the previous AFM imaging study (13) are in reasonable agreement: in this condition, collagen is well described by the standard WLC model and curvature may be disregarded. In general, however, our findings demonstrate the importance of mapping the functional forms of $\langle R^2(s) \rangle$ and $\langle \cos\theta(s) \rangle$, and caution against determining persistence length by using only chain end-point separations.

In our description and analysis of collagen, we have made two major assumptions. First, neither the WLC nor cWLC model takes into account the helical nature of collagen. Thus, flexibility of the chain derives only from bending deformations. If the intrinsic curvature described by our model arises due to a surface-collagen interaction inducing altered twist in the protein, then torsional stiffness of the triple helix (50) is an important parameter, and twist-bend coupling (51) may be required to fully describe collagen's flexibility.

Second, our analysis procedures treat collagen as an apolar, homogenous worm-like chain. We are not able to discriminate between N→C and C→N directionality in the imaged chains. Furthermore, persistence lengths and curvatures reported in this work represent average values for the entire collagen sequence. In adopting this approach, we have implicitly assumed that the (Gly-X-Y)$_n$ triple-helix-forming motif dominates collagen's response, with local sequence variations (e.g. fraction of prolines occupying the X and Y positions) viewed as introducing only minor perturbations. The flexibility of fibrillar collagen triple helices appears approximately uniform in EM images (18), supporting this assumption. However, molecular dynamics simulations (52) and modelling of atomistic collagen structures (50) suggest that collagen possesses a sequence-dependent flexibility, commensurate with the expected variations in pitch observed for different triple helical sequences (2, 40). It is interesting to speculate how a sequence-dependent mechanical signature may provide additional physical cues to interaction partners (44). Experimentally elucidating the detailed sequence-dependent flexibility of collagen is an exciting future prospect that will require higher-precision algorithms than applied thus far to this mechanically important protein.

**CONCLUSIONS AND OUTLOOK**

Using atomic force microscopy, we determined the persistence length of different types and sources of collagen in the presence of different co-solutes ($K^+$, $Cl^-$, acetate). Although collagens of different type and source exhibited only minor differences in persistence length, our initial findings implied that ionic strength and, to a lesser extent, pH modulate collagen's flexibility. Closer inspection of these results, however, revealed this interpretation to be an artefact of the model used for analysis. Rather than behaving as an intrinsically straight worm-like chain, collagen deposited in low salt conditions was found to adopt preferentially curved conformations on the mica surface. Statistical analysis of chain conformations demonstrates that this curvature is a thermodynamic property, with chains appearing equilibrated in two dimensions. The extent of curvature depends strongly on salt concentration and pH, with persistence length remaining roughly constant across all conditions. Thus, collagen possesses an experimentally tunable curvature, and is to our knowledge the first such example of a curved worm-like chain.

Our findings provide a possible explanation for the range of persistence lengths reported for collagen in the literature: rather than a direct modulation of its flexibility, the presence of certain co-solutes may induce curvature along the collagen backbone. Whether or not the curvature seen in our results is intrinsic to collagen or is caused by co-solute-dependent interactions with the imaging surface remains to be determined. If intrinsic, then the modulation of collagen's conformation by salt and pH has potentially broad implications. The acidification experienced along the secretory pathway (24) may play a heretofore unexplored role in controlling collagen's intracellular compactness. Changes in salt concentration experienced upon secretion may also contribute to extracellular self-assembly of fibrillar collagens; for example, chloride ions are essential for the distinct network assembly of type IV collagen (53). Such changes in pH and co-solute concentrations could therefore influence collagen self-assembly through modification of its conformational properties: collagen does not assemble into fibrillar structures at low pH or at low ionic strength (35, 54), precisely those conditions in which we observe its strongest curvature.

The strong agreement between measures of chain flexibility and the predictions of the curved worm-like chain model suggest that this formalism may be useful for interpreting the mechanical properties of other biological filaments. Examples such as amyloid fibrils (33) and coiled-coil proteins (55) have been found to exhibit preferentially curved conformations at interfaces; applying the cWLC model to describe these systems and others is likely to provide further insight into their equilibrium properties and the energies governing their mechanical properties.

**Table 2** – Summary of fitting parameters determined in this work.

| Collagen Type | Solvent Condition | Standard Model Fits | | Curved Model Fits | | | |
|---|---|---|---|---|---|---|---|
| | | $p$ (nm) | $\Delta p$ (nm) | $p$ (nm) | $\Delta p$ (nm) | $\kappa_o$ (nm$^{-1}$) | $\Delta\kappa_o$ (nm$^{-1}$) |
| Rat Type-I | Water | 35 | 8 | 108 | 22 | 0.015 | 0.002 |
| Rat Type-I | 100 µM KCl | 30 | 7 | 76 | 18 | 0.016 | 0.002 |
| Rat Type-I | 1 mM KCl | 48 | 7 | 102 | 18 | 0.0114 | 0.0009 |
| Rat Type-I | 10 mM KCl | 93 | 6 | 101 | 14 | 0.002 | 0.005 |
| Rat Type-I | 100 mM KCl | 145 | 13 | 144 | 41 | 0 | N/A |
| Rat Type-I | 1 mM HCl | 28 | 4 | 49 | 7 | 0.015 | 0.001 |
| Rat Type-I | 10 mM KCl + 1 mM HCl | 54 | 7 | 131 | 33 | 0.011 | 0.001 |
| Rat Type-I | 100 mM KCl + 1 mM HCl | 112 | 4 | 117 | 14 | 0.002 | 0.002 |
| Human Type-I | 100 mM KCl + 1 mM HCl | 91 | 4 | 98 | 11 | 0.002 | 0.002 |
| Human Type-II | 100 mM KCl + 1 mM HCl | 94 | 4 | 96 | 14 | 0.0008 | N/A |
| Human Type-III | 100 mM KCl + 1 mM HCl | 85 | 5 | 90 | 15 | 0.0018 | N/A |
| Rat Type-I | 20 mM Acetic Acid | 23 | 4 | 51 | 7 | 0.018 | 0.002 |
| Human Type-I | 20 mM Acetic Acid | 26 | 4 | 51 | 4 | 0.0166 | 0.0006 |
| Human Type-II | 20 mM Acetic Acid | 31 | 7 | 76 | 13 | 0.0158 | 0.0008 |
| Human Type-III | 20 mM Acetic Acid | 28 | 5 | 60 | 4 | 0.0163 | 0.0005 |

Fit parameters are averages of the values from independent $\langle R^2(s)\rangle$ and $\langle \cos\theta(s)\rangle$ fits of the data; errors, $\Delta$, represent the error in these estimates. The total contour lengths of collagen traced in each condition is provided along with a comprehensive list of fitting parameters and $\chi^2$ values in Table S1.

**Data availability**

The datasets generated and analysed during the current study are available from the corresponding author on request.

**SUPPORTING MATERIAL**

Supporting Methods, Supporting Table 1 and Supporting Figures S1-S7 are available in the supporting material.

**AUTHOR CONTRIBUTIONS**

NR and NRF designed the project. NR and AL performed experiments. NR developed the SmarTrace algorithm and software. AL performed polymer chain simulations and derived the cWLC fitting equations. NR and AL analysed the data. All authors contributed to the writing of the manuscript.

**ACKNOWLEDGMENTS**

This work was supported by a Discovery Grant from the Natural Sciences and Engineering Research Council of Canada (NSERC). We thank Alex Dunn for collagen samples, Alaa Al-Shaer for testing robustness of the SmarTrace algorithm, Peter Olmsted for introducing to us the concept of surface-induced fiber curvature, David Sivak for a critical review of the manuscript, the Forde lab for many useful discussions, ChangMin Kim for technical support, and Guillaume Lamour for the EasyWorm code, DNA sample images and an introduction to AFM sample preparation.

**SUPPORTING CITATIONS**

References 56-59 appear in the supporting material.

# Environmentally controlled curvature of single collagen proteins

# Supporting Information

Naghmeh Rezaei, Aaron Lyons and Nancy R. Forde*
Department of Physics
Simon Fraser University
8888 University Drive
Burnaby, BC  V5A 1S6
CANADA

*Corresponding author.  778-782-3161.  nforde@sfu.ca.  ORCID: 0000-0002-5479-7073

## Derivation of the curved Worm-Like Chain (cWLC) model

The standard WLC model assumes that there is an energetic cost to bend an intrinsically straight rod. Specifically, to bend a semi-flexible, intrinsically straight rod with length $\delta l$ into a circular arc with central angle $\delta\theta$, the energy required is (1)

$$E_{\text{bend}} = \frac{\alpha \delta\theta^2}{2\delta l} = \frac{p\delta\theta^2}{2\delta l} k_B T. \tag{S1}$$

$\alpha$ is the bending rigidity of the chain, related to the persistence length $p$ through the relation $p = \alpha/k_B T$, where $k_B T$ is the product of the Boltzmann constant and the absolute temperature. The central angle of this arc, $\delta\theta$, is equivalent to the angle between the tangents at the beginning and end of the segment. The lowest energy conformation of the segment is a straight line, with the energy increasing harmonically about this configuration. The Boltzmann distribution provides the distribution of angles that this segment adopts in the presence of thermal noise, which is given by

$$P(\delta\theta) = \sqrt{\frac{p}{2\pi\delta l}} \exp\left(-\frac{p(\delta\theta)^2}{2\delta l}\right). \tag{S2}$$

This a normal distribution with mean $\langle\delta\theta\rangle = 0$ and variance $\sigma^2_{\delta\theta} = \frac{\delta l}{p}$.

If the chain is intrinsically curved, the bending energy is modified to reflect deviations from this bent state.

$$E_{\text{bend}} = \frac{p(\delta\theta - \kappa_o \delta l)^2}{2\delta l} k_B T, \tag{S3}$$

where $\kappa_o$ is the intrinsic curvature of the chain, defined as $\kappa_o = \langle\frac{d\theta}{dl}\rangle$. The distribution of angles that this intrinsically curved segment adopts in the presence of thermal noise is given by

$$P(\delta\theta) = \sqrt{\frac{p}{2\pi\delta l}} \exp\left(-\frac{p(\delta\theta - \kappa_o \delta l)^2}{2\delta l}\right). \tag{S4}$$

As before, this is a normal distribution with variance $\sigma^2_{\delta\theta} = \frac{\delta l}{p}$, but now centered about a mean $\langle\delta\theta\rangle = \kappa_o \delta l$.

We now consider a longer chain segment, comprised of $N$ segments each of length $\delta l$, such that the total length of this chain $l = N\delta l$. This longer chain is not necessarily a circular arc, but is made up of $N$ short circular arc segments, each with angular difference $\delta\theta_i$ distributed about $\kappa_o \delta l$ as per equation (S4). The angular difference $\theta$ between the ends of this larger segment is given by the sum of the $N$ angles adopted by the smaller circular arcs:

$$\theta = \sum_{i=1}^{N} \delta\theta_i. \tag{S5}$$

Since the sum of normal random variables is normally distributed, the full distribution of $\theta$ is completely described by its mean and variance, given respectively by

$$\langle \theta \rangle = N\langle \delta\theta \rangle = \kappa_o N \delta l = \kappa_o l \tag{S6}$$

and

$$\sigma_\theta^2 = N\sigma_{\delta\theta}^2 = \frac{N\delta l}{p} = \frac{l}{p}. \tag{S7}$$

The angular distribution of this segment of length $l$ is therefore given by

$$P(\theta) = \sqrt{\frac{p}{2\pi l}} e^{-\frac{p(\theta - \kappa_o l)^2}{2l}}. \tag{S8}$$

Note that, because $\langle \theta \rangle \neq 0$, this chain exhibits a net global curvature.

For any arbitrary segment length $s$, we can now calculate the average tangent vector correlation as

$$\langle \hat{t}(s+s') \cdot \hat{t}(s') \rangle = \langle \cos\theta(s) \rangle = \sqrt{\frac{p}{2\pi l}} \int_{-\infty}^{\infty} \exp\left(-\frac{p(\theta - \kappa_o s)^2}{2s}\right) \cos\theta \, d\theta, \tag{S9}$$

where $s'$ defines an arbitrary starting position of the segment along the contour. This simplifies to

$$\langle \cos\theta \rangle = \exp\left(-\frac{s}{2p}\right) \cos(\kappa_o s). \tag{S10}$$

However, since $\hat{t}(s+s') \cdot \hat{t}(s') \leq 1$ always, then its average $\langle \hat{t}(s+s') \cdot \hat{t}(s') \rangle \leq 1$. This requires that $s \geq 0$, so we can write

$$\langle \cos\theta(s) \rangle = \exp\left(-\frac{|s|}{2p}\right) \cos(\kappa_o |s|) \tag{S11}$$

for all $s$.

Similarly, we can calculate the mean squared end-to-end distance of the segment as

$$\langle R^2(s) \rangle = \langle \vec{R}(s) \cdot \vec{R}(s) \rangle = \left\langle \left(\int_0^s \hat{t}(s')ds'\right)\left(\int_0^s \hat{t}(s'')ds''\right) \right\rangle = \int_0^s \int_0^s \langle \hat{t}(s') \cdot \hat{t}(s'') \rangle ds'ds'', \tag{S12}$$

where the order of operations was interchanged because averaging and integrating are both linear operations. Using $\langle \hat{t}(s') \cdot \hat{t}(s'') \rangle = \exp\left(-\frac{|s''-s'|}{2p}\right) \cos(\kappa_o |s'' - s'|)$ from equation (S11) gives

$$\langle R^2(s) \rangle = \int_0^s \left\{ \int_0^{s''} \exp\left[-\frac{(s''-s')}{2p}\right] \cos[\kappa_o(s''-s')] \, ds' + \int_{s''}^s \exp\left[-\frac{(s'-s'')}{2p}\right] \cos[\kappa_o(s'-s'')] \, ds' \right\} ds''. \quad (S13)$$

Evaluating this expression yields

$$\langle R^2(s) \rangle = \frac{4sp}{(1+4\kappa_0^2 p^2)^2} \left\{ 1 - \frac{2p}{s}(1 - 4\kappa_0^2 p^2)\left[1 - \cos(\kappa_0 s) \exp\left(-\frac{s}{2p}\right)\right] + \frac{4\kappa_0 p^2}{s}\left[\kappa_0 s - 2\sin(\kappa_0 s) \exp\left(-\frac{s}{2p}\right)\right] \right\}. \quad (S14)$$

Once again, this expression is valid only for $s \geq 0$, as negative values of $s$ can produce values of $\langle R^2(s) \rangle$ that are negative. Thus,

$$\langle R^2(s) \rangle = \frac{4|s|p}{(1+4\kappa_0^2 p^2)^2} \left\{ 1 - \frac{2p}{|s|}(1 - 4\kappa_0^2 p^2)\left[1 - \cos(\kappa_0 |s|) e^{-\frac{|s|}{2p}}\right] + \frac{4\kappa_0 p^2}{|s|}\left[\kappa_0 |s| - 2\sin(\kappa_0 |s|) e^{-\frac{|s|}{2p}}\right] \right\}. \quad (S15)$$

In the case where $\kappa_o = 0$, which corresponds to the standard two-dimensional worm-like chain, the expressions for the tangent vector correlation and mean squared end-to-end distance reduce to

$$\langle \cos\theta(s) \rangle = \exp\left(-\frac{|s|}{2p}\right) \quad (S16)$$

and

$$\langle R^2(s) \rangle = 4|s|p\left\{1 - \frac{2p}{|s|}\left[1 - \exp\left(-\frac{|s|}{2p}\right)\right]\right\}. \quad (S17)$$

These are identical to previously derived results (2), and to equations (2) and (1) in the main text, respectively.

**SmarTrace Algorithm**

From AFM images of biopolymers, we wished to analyze chain configurations and extract mechanical properties. For this purpose, we developed a robust and efficient chain tracing software in MATLAB, dubbed "SmarTrace", which traces the centerline of each molecule with sub-pixel resolution and provides comprehensive statistical analysis of the chains.

Chain tracing

An overview of the SmarTrace workflow is discussed below. Details are available in reference (3).

1. In a visual user interface (adapted from the EasyWorm package (4) within MATLAB (5)), the user selects a few points on or near the backbone of the chain to be traced.
2. SmarTrace fits an initial spline to these user-defined points and extracts points along the spline separated by one nanometer.
3. The program uses top-hat and median filtering (6) to improve the signal-to-noise ratio of the region surrounding the chain.
4. To detect the best path describing the chain, a search window is defined for each point on the spline curve along the tangential direction of the initial spline. A search grid with sub-pixel resolution determines (interpolated) intensity values of the image for each grid point.
5. For each point on the initial spline, a template pattern is matched with each point in the grid within the search window. The template resembles the intensity pattern of a cross-section of the chain with varying widths.
6. A matching score is calculated for each possible width and location of this pattern. Cross-correlation scores are used to determine the best centerline position and width of the chain.
7. To ensure stable results, a penalty term is added for sudden changes in width and/or direction of the chain.
8. After the scores are finalized, the re-weighted maximum scores are used to extract the width and center of the chain at points spaced approximately 1 nanometer apart. A B-spline is fit to these points, resulting in a piecewise smooth polynomial that represents the polymer chain.

This method is not very sensitive to image quality and can successfully detect the chain backbone in noisy images; even if parts of the chain are slightly faded, the code is still able to trace the entire chain. The results also do not depend on where the user selects the points (validated by having different users trace the same set of experimental chains, and finding the same persistence length), and the initially selected points do not need to be located exactly on the chain centerline. These features, along with an efficient computation algorithm, make the code fast and easy to use for tracing and analyzing images of single polymers, as obtained for example by atomic force microscopy, electron microscopy or fluorescence microscopy.

Statistical analysis of chain properties

After the chains have been traced, statistical approaches are used to analyze their flexibility. First, the traced chains are partitioned into segments of varying lengths. Utilizing a method introduced in reference (7), each molecule is divided into multiple segments, with lengths drawn randomly from a pre-defined set of input values (here, 10 nm, 20 nm, 30 nm, ..., 200 nm). Once the lengths of these segments have been determined, their positions on the chain are shuffled to avoid accumulating shorter lengths towards one end of the chain. This process is repeated 50 times for each chain, allowing different regions of the chain to contribute to the statistics of different segment lengths. Within each draw, the sampled chain segments are nonoverlapping. As rare, longer segments are less useful for mathematical fitting and persistence length calculations, choosing a relatively short maximum segment length provides more samples in each bin. This method also allows for the use of partially traced chains, which is particularly helpful when ends of a molecule are not clear or chains intersect.

Several statistics are then calculated from the samples at each segment length $s$: the mean-squared end-to-end distance, $\langle R^2(s) \rangle$, and the mean cosine of the angle between the start and end of the segment, $\langle \cos \theta(s) \rangle$.

**Validation of SmarTrace and analysis procedures**

Validation with DNA images

Images of DNA molecules with 1 μm contour length, provided with the software package Easyworm (4), were traced and analyzed with SmarTrace. A total of 24 molecules was used in the analysis. The mean squared end-to-end distance, $\langle R^2(s) \rangle$, is shown in Figure S1A along with a 2D worm-like chain fit (Equation 1). Figure S1B shows the experimental data for tangent vector correlation, $\langle \cos \theta(s) \rangle$, and the 2D WLC fit (Equation 2). A persistence length of 52 ± 2 nm was obtained from these analyses, consistent with expectations for DNA (2, 4).

Validation with simulated worm-like chains

To validate the methodology used for chain tracing and analysis within SmarTrace, two-dimensional worm-like chains were simulated and converted into pseudo-AFM images. These images were then traced with SmarTrace to ensure that the algorithm was able to accurately recover the input persistence length and curvature. The workflow for the simulation is as follows:

1. The desired persistence length $p$, curvature $\kappa_0$, contour length $L$ and width of the chains $w$ – as well as the average number of chains per image – are input by the user.
2. For the first chain, an angle $\theta_0$ between 0 and 360º, as well as two values $x_0$ and $y_0$ between 0 and 2000 nm are sampled randomly from a uniform distribution. These values represent the starting angle and initial $xy$-coordinates of the chain, respectively.
3. The curvature of the chain is chosen to be either $\kappa_0$ or $-\kappa_0$ with equal chance; this represents the ability of the chain to "lie down" on the mica surface with either a left-handed or right-handed curvature.

4. Using a step size $\delta s = 0.5$ nm, an angle $\delta\theta$ is sampled from a normal distribution with mean $\pm\kappa_o \delta s$ and variance $\frac{\delta s}{p}$.
5. The next point on the chain is placed at $x_1 = x_0 + \delta s \cos(\theta_1)$ and $y_1 = y_0 + \delta s \sin(\theta_1)$, where $\theta_1 = \theta_0 + \delta\theta$.
6. This process is repeated, choosing a random $\delta\theta$ as above, generating $\theta_{i+1} = \theta_i + \delta\theta$ and placing the next point on the chain at $x_{i+1} = x_i + \delta s \cos(\theta_{i+1})$ and $y_{i+1} + \delta s \sin(\theta_{i+1})$.
7. This process is terminated after $n$ steps, where $n\delta s = L$.
8. Steps 2 through 7 are repeated for every chain in the image.
9. A 2000-by-2000 array of pixels is overlaid on top of the image, discretizing the image such that each pixel represents 1 nm².
10. The intensity of each pixel is populated by considering every point on each chain as an intensity source, which contributes an intensity

$$I(x,y) = I_o \exp\left(-\frac{(x-x_i)^2 + (y-y_i)^2}{w}\right) \quad \text{(S18)}$$

to the pixel centered at $(x, y)$. The intensity contributions from every point on every chain are then summed to yield the overall intensity of this pixel.
11. Once the intensities of all pixels have been computed, the image is converted into a 500-by-500 array of pixels by averaging the intensities of 4-by-4 pixel blocks. The intensities of these now 4x4 nm² pixels (the pixel size of our experimental images) are then normalized to the maximum intensity pixel in the image, and converted into an 8-bit grayscale image.
12. The chains detailed in these images can then be traced with SmarTrace to extract their persistence length and curvature.

We generated two sets of images with parameters chosen to emulate experimentally gathered AFM images of collagen: both sets were given a width parameter of 4 nm and a contour length of 300 nm – chosen to replicate AFM images of collagen. For simplicity, background noise was not included in these simulations.

The first set was generated with a persistence length of 85 nm and zero curvature. An example image and standard WLC model fits to the data (Equations 1, 2 from the main text) are shown in Figures S2A-C. The traced data also reproduce the Gaussian properties of the simulated chains, as shown by the kurtosis and normality of the angular distributions at different segment lengths (Figures S2D and S2E). The standard error in the kurtosis (SEK), used to generate the error bars in Figure S2D, is given by (8)

$$\text{SEK} = \sqrt{\frac{24n(n-1)^2}{(n-3)(n-2)(n+3)(n+5)}}, \quad \text{(S19)}$$

where $n$ is the number of observations comprising the distribution. The expected distribution shown in Figure S2E is given by Eq. S8 with $p = 85$ nm, $s = 50$ nm and $\kappa_o = 0$ nm$^{-1}$.

The second set of chains was generated with a persistence length of 50 nm and a curvature of 0.02 nm$^{-1}$; an image of these chains is shown in Figure S3A. Due to the presence of chains with both positive and negative curvature (see step 3, above), the angular distributions extracted from these curved chains will be a sum of two Gaussian distributions, with mean $\pm\kappa_o s$ and variance $\frac{s}{p}$. These two distributions may not have equal amplitude (for a small number of sampled chains), making the properties of the expected distribution difficult to calculate. However, the expected distribution of the absolute-values of the angles is given by

$$P(|\theta|) = \sqrt{\frac{p}{2\pi s}} \left[ \exp\left(-\frac{p(|\theta|-\kappa_o s)^2}{2s}\right) + \exp\left(-\frac{p(|\theta|+\kappa_o s)^2}{2s}\right) \right] \quad (S20)$$

regardless of the proportion of the two distributions. Figure S3B shows the expected distribution for $p = 50$ nm, $\kappa_o = 0.02$ nm$^{-1}$ and $s = 50$ nm, plotted with a histogram of the angles extracted from the simulated chains at a 50 nm segment length, demonstrating good agreement between the two distributions. Figures S3C and S3D show fits of the curved WLC expressions (Equations 3, 4) to the traced data, both of which yield persistence lengths and curvatures close to the input parameters.

| Collagen Type | Solution | Length Traced (μm) | Standard Model Fits | | | | | | Curved Model Fits | | | | | | | | | |
|---|---|---|---|---|---|---|---|---|---|---|---|---|---|---|---|---|---|---|
| | | | $\langle R^2 \rangle$ | | | $\langle \cos\theta \rangle$ | | | $\langle R^2 \rangle$ | | | | | $\langle \cos\theta \rangle$ | | | | |
| | | | $p$ | $\Delta p$ | $\chi^2_{red}$ | $p$ | $\Delta p$ | $\chi^2_{red}$ | $p$ | $\Delta p$ | $\kappa_o$ | $\Delta\kappa_o$ | $\chi^2_{red}$ | $p$ | $\Delta p$ | $\kappa_o$ | $\Delta\kappa_o$ | $\chi^2_{red}$ |
| Rat Type-I | Water | 14.7 | 32 | 4 | 687 | 38 | 10 | 101 | 130 | 29 | 0.017 | 0.0004 | 4.1 | 87 | 9 | 0.0126 | 0.0004 | 1.7 |
| Rat Type-I | 100 μM KCl | 6.3 | 28 | 3 | 88 | 32 | 9 | 24 | 76 | 21 | 0.0173 | 0.0009 | 12 | 75 | 13 | 0.0142 | 0.001 | 9.4 |
| Rat Type-I | 1 mM KCl | 12.4 | 49 | 4 | 191 | 46 | 9 | 60 | 116 | 23 | 0.0123 | 0.0007 | 2.3 | 89 | 12 | 0.0105 | 0.0006 | 1.7 |
| Rat Type-I | 10 mM KCl | 17.2 | 98 | 3 | 5.6 | 87 | 4 | 2.0 | 115 | 13 | 0.0044 | 0.0014 | 12 | 88 | 8 | 0.0007 | 0.006 | 2.2 |
| Rat Type-I | 100 mM KCl | 19.9 | 151 | 8 | 47 | 139 | 16 | 28 | 149 | 39 | 0 | N/A | 47 | 139 | 42 | 0 | N/A | 30 |
| Rat Type-I | 1 mM HCl | 20.4 | 27 | 2 | 423 | 29 | 6 | 68 | 52 | 9 | 0.0163 | 0.001 | 18 | 47 | 6 | 0.015 | 0.001 | 4.4 |
| Rat Type-I | 10 mM KCl + 1 mM HCl | 33.6 | 55 | 5 | 299 | 52 | 9 | 101 | 164 | 25 | 0.0123 | 0.0004 | 7.7 | 98 | 7 | 0.0095 | 0.0003 | 1.2 |
| Rat Type-I | 100 mM KCl + 1 mM HCl | 20.9 | 116 | 3 | 20 | 107 | 6 | 14 | 122 | 14 | 0.002 | 0.003 | 27 | 113 | 14 | 0.0022 | 0.002 | 18 |
| Human Type-I | 100 mM KCl + 1 mM HCl | 20.7 | 90 | 3 | 7.7 | 92 | 4 | 11 | 106 | 11 | 0.005 | 0.001 | 21 | 91 | 10 | 0 | N/A | 11 |
| Human Type-II | 100 mM KCl + 1 mM HCl | 12 | 93 | 3 | 5.2 | 96 | 5 | 2.2 | 93 | 15 | 0 | N/A | 5.4 | 98 | 12 | 0.002 | 0.003 | 2.4 |
| Human Type-III | 100 mM KCl + 1 mM HCl | 16.7 | 85 | 3 | 2.1 | 84 | 6 | 8.6 | 85 | 15 | 0 | N/A | 2.2 | 95 | 14 | 0.004 | 0.002 | 12 |
| Rat Type-I | 20 mM Acetic Acid | 24.7 | 22 | 2 | 501 | 25 | 5 | 62 | 58 | 6 | 0.0200 | 0.0004 | 3.7 | 44 | 3 | 0.0163 | 0.0008 | 2.0 |
| Human Type-I | 20 mM Acetic Acid | 50.4 | 26 | 2 | 788 | 26 | 6 | 228 | 52 | 6 | 0.0167 | 0.0007 | 33 | 49 | 2 | 0.0165 | 0.0004 | 3.1 |
| Human Type-II | 20 mM Acetic Acid | 13.2 | 29 | 3 | 865 | 33 | 10 | 170 | 74 | 17 | 0.0165 | 0.0008 | 35 | 78 | 6 | 0.0150 | 0.0005 | 2.1 |
| Human Type-III | 20 mM Acetic Acid | 40.4 | 28 | 2 | 1248 | 28 | 7 | 253 | 64 | 5 | 0.0168 | 0.0003 | 31 | 56 | 3 | 0.0158 | 0.0005 | 3.0 |

**Table S1** - Comprehensive list of fitting parameters obtained for different collagen types and co-solute conditions. Persistence lengths, $p$, are given in nm, while curvature $\kappa_0$ is in units of nm$^{-1}$. Reported errors, $\Delta$, represent 95% confidence intervals in the fitting parameters.

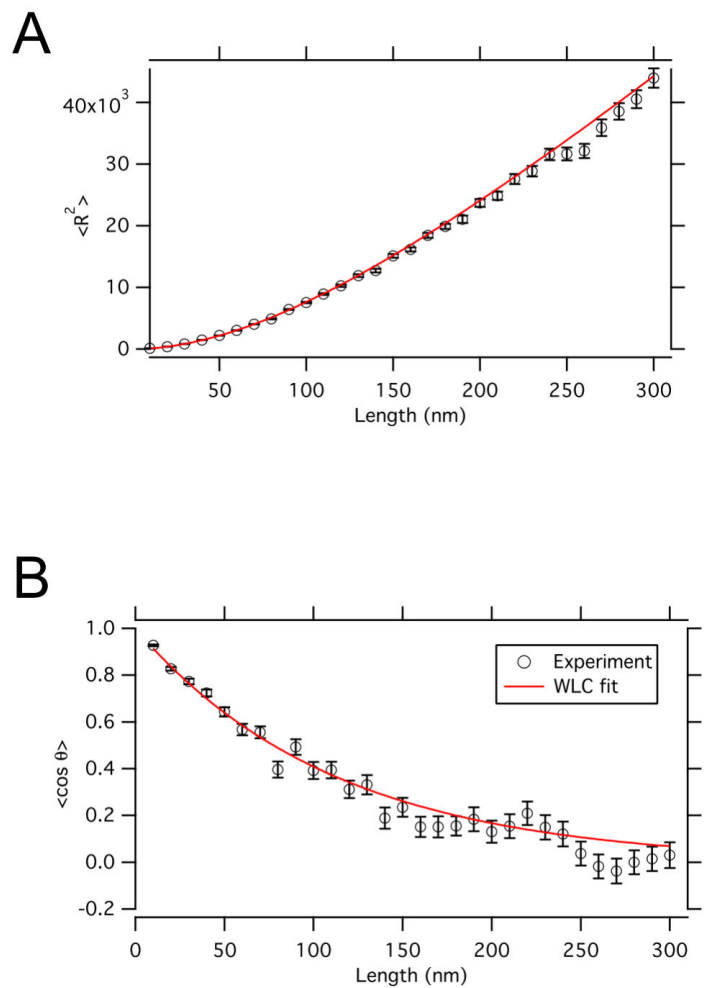

**Figure S1.** (A). Mean squared end-to-end distance versus length along DNA molecules. Symbols represent the mean for each segment length while error bars show the standard error of the mean. The persistence length from the fit (red line; Equation 1) is 52 ± 1 nm. (B) Average tangent-tangent correlation vs. length along DNA molecules. The persistence length from the fit (red line; Equation 2) is 52 ± 3 nm.

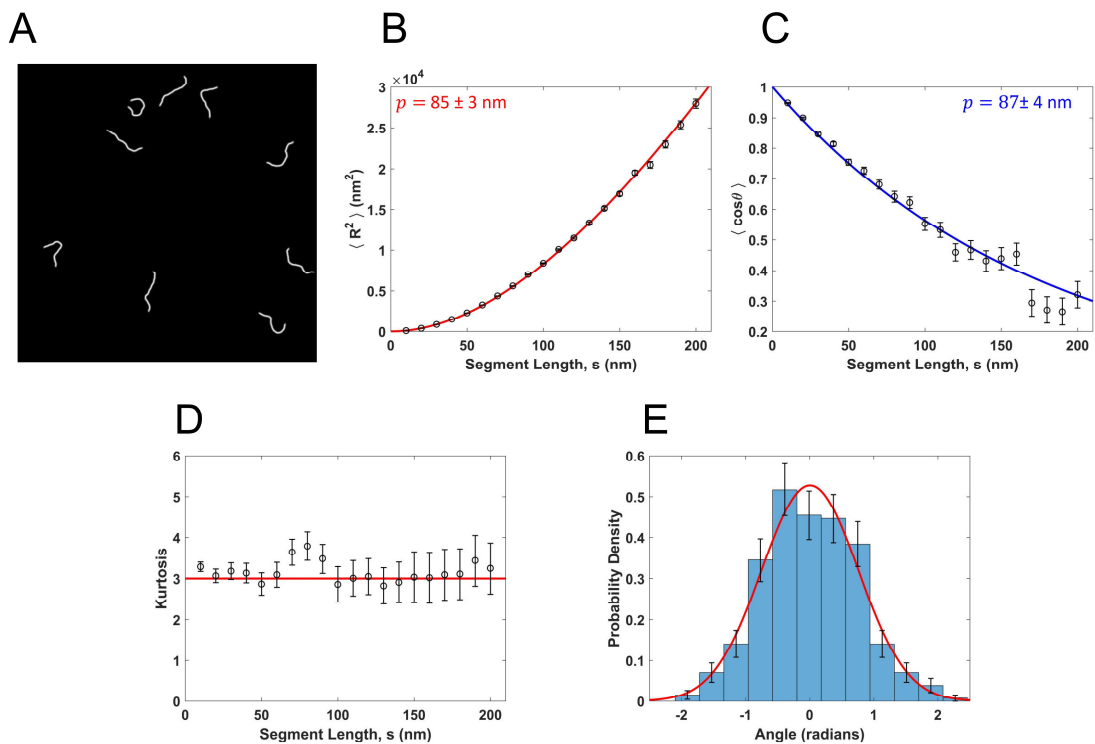

**Figure S2**. Test of SmarTrace analysis procedures using simulated worm-like chains. (A) Example image of simulated worm-like chains with a persistence length of $p = 85$ nm. After being traced, the data were fit with the standard WLC expressions for mean squared end-to-end distance (Eq. 1) and mean tangent vector correlation (Eq. 2), as shown in (B) and (C), respectively. Both fits yield persistence lengths consistent with the value input into the simulations. (D) Kurtosis of the angular distributions extracted from the simulated chains at different segment lengths. A kurtosis near the expected value of 3 is achieved at all length scales. Errors in the kurtosis are given by Eq. S19. (E) Individual angular distributions are also well described as Gaussian, as shown by good agreement between the expected and extracted angular distributions at a 50 nm segment length.

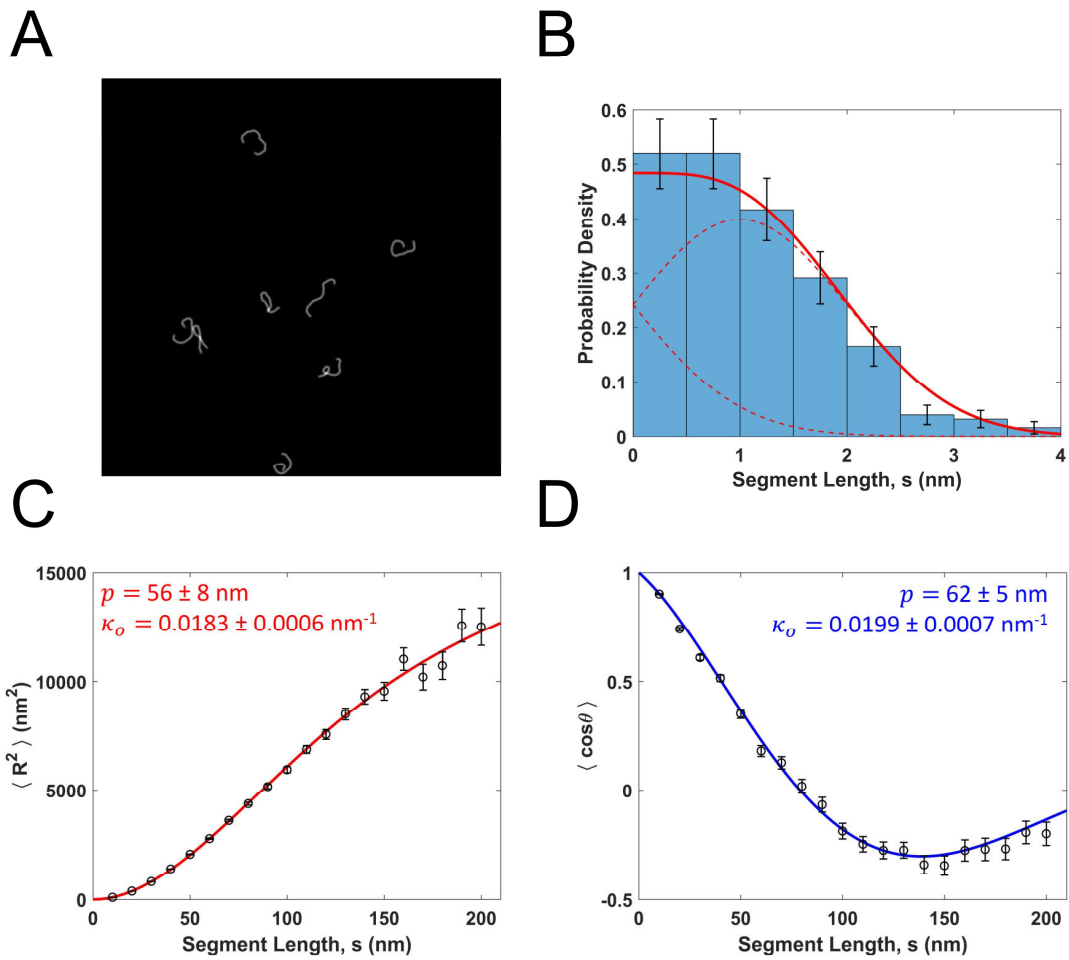

**Figure S3**. Analysis of simulated curved worm-like chains. (A) Example image of simulated curved worm-like chains with a persistence length of $p = 50$ nm and curvature of $\kappa_o = 0.02$ nm$^{-1}$. (B) Histogram of the absolute-valued angles extracted from the traces of these simulated chains at a segment length of $s = 50$ nm, as well as the expected probability distribution (solid red line, Eq. S20) and the two Gaussian components that comprise it. (C, D) Fits of data from these simulated chains to the curved worm-like chain expectations for mean squared end-to-end distance (Eq. 3) and mean tangent vector correlation (Eq. 4), respectively. Both fits yield parameters close to those input in the chain simulations.

- 12 -

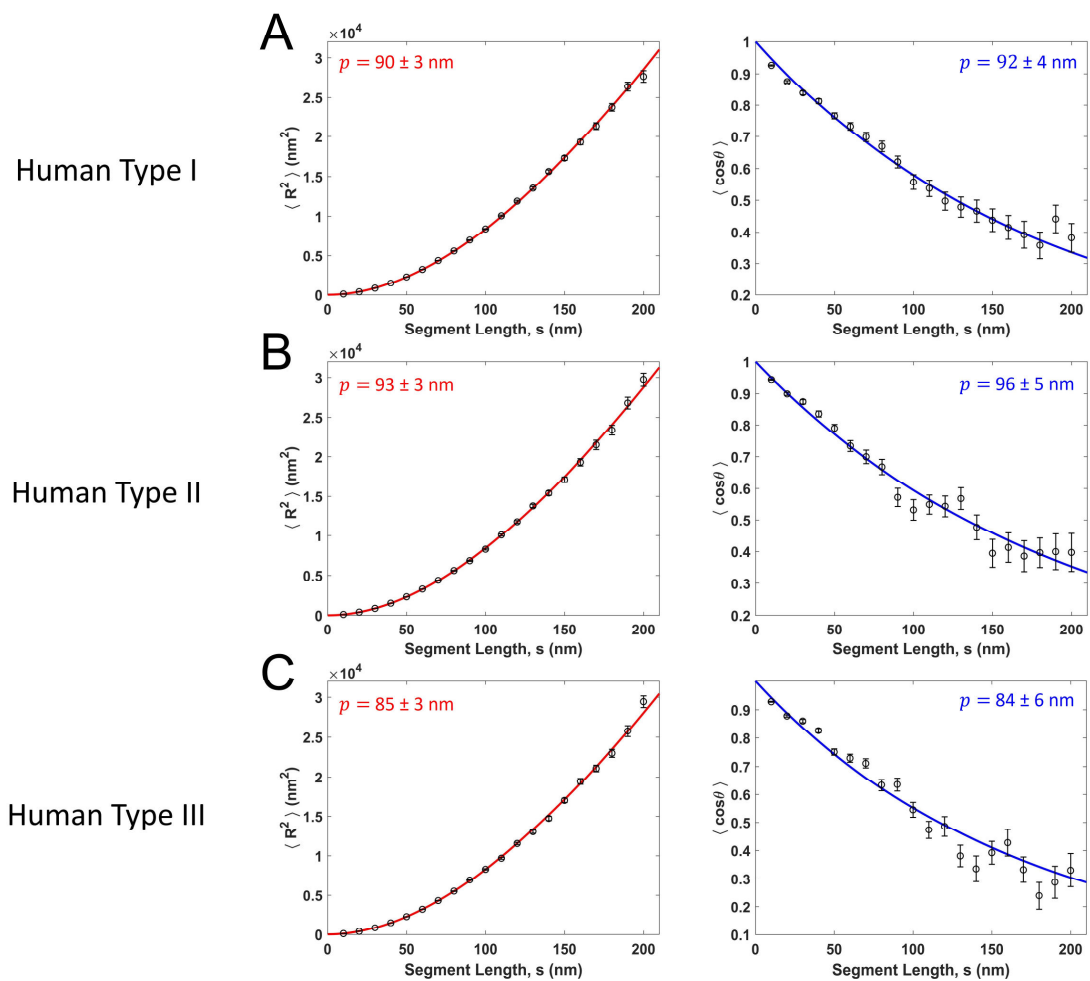

**Figure S4.** Standard WLC model fits to human collagens (A) type I, (B) type II, and (C) type III, all deposited from 100 mM KCl with 1 mM HCl. ⟨$R^2$⟩ fits (Eq. 1) are shown in the left column in red, and ⟨$\cos\theta$⟩ fits (Eq. 2) are shown in the right column in blue.

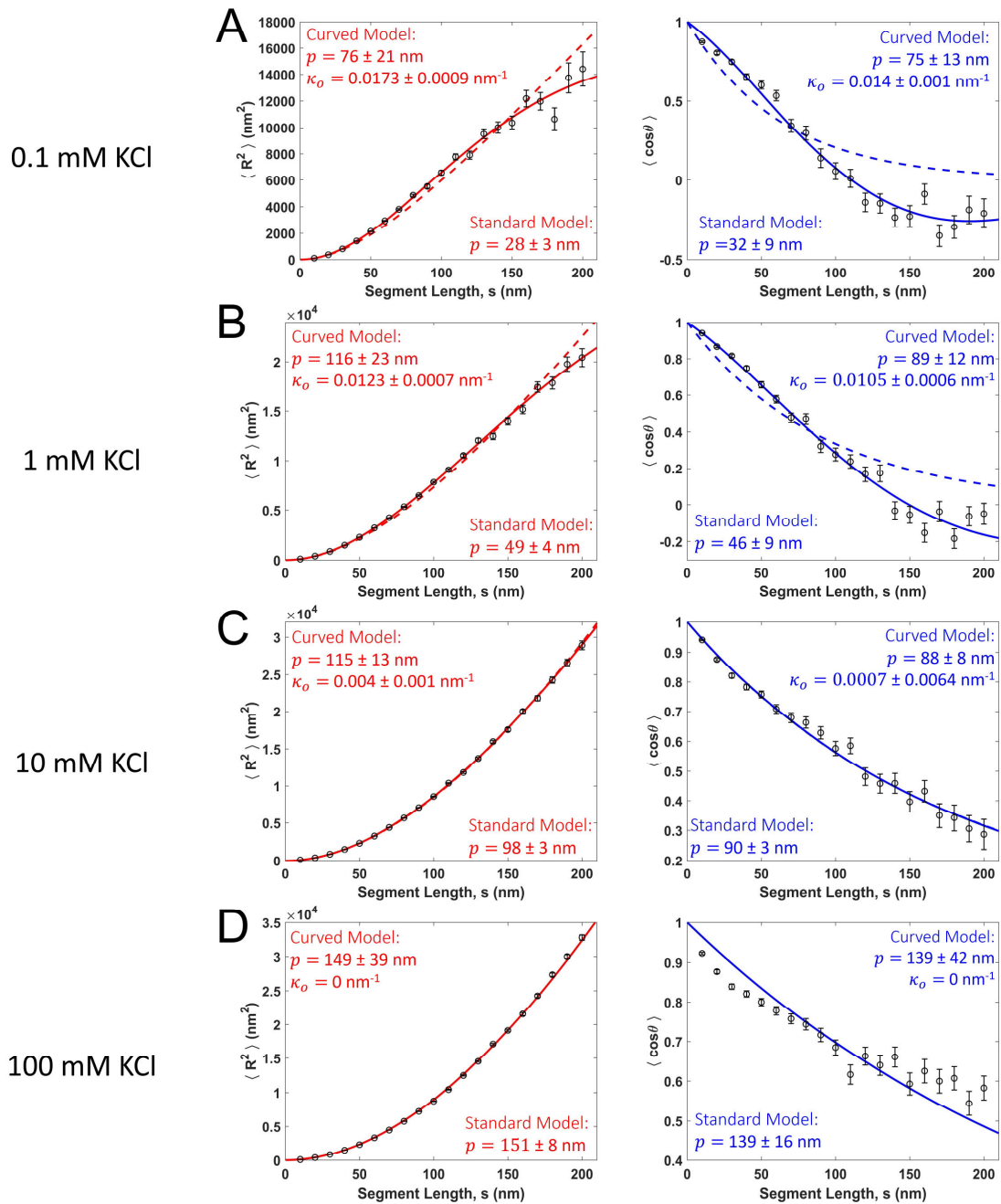

**Figure S5.** Standard (dashed lines) and curved (solid lines) WLC model fits to rat tail collagen type I data deposited from (A) 0.1 mM KCl, (B) 1 mM KCl, (C) 10 mM KCl and (D) 100 mM KCl. $\langle R^2 \rangle$ fits are shown in the left column in red, and $\langle \cos\theta \rangle$ fits are shown in the right column in blue. As the KCl concentration is increased, the standard and curved fits converge to the same result.

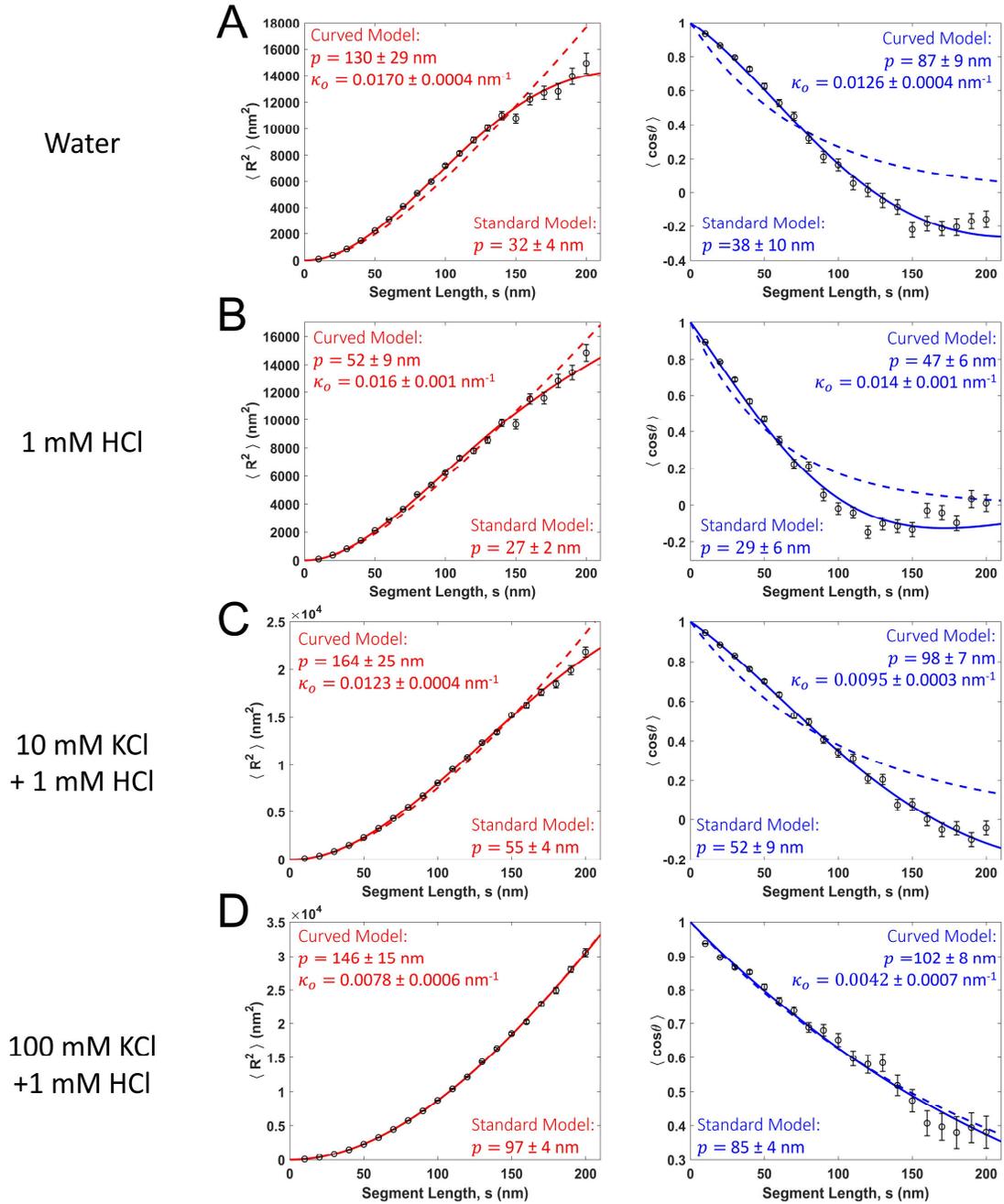

**Figure S6.** Standard (dashed lines) and curved (solid lines) WLC model fits to rat tail collagen type I data from (A) water, (B) 10 mM KCl + 1 mM HCl and (C) 100 mM KCl + 1 mM HCl. $\langle R^2 \rangle$ fits are shown in the left column in red, and $\langle \cos\theta \rangle$ fits are shown in the right column in blue.

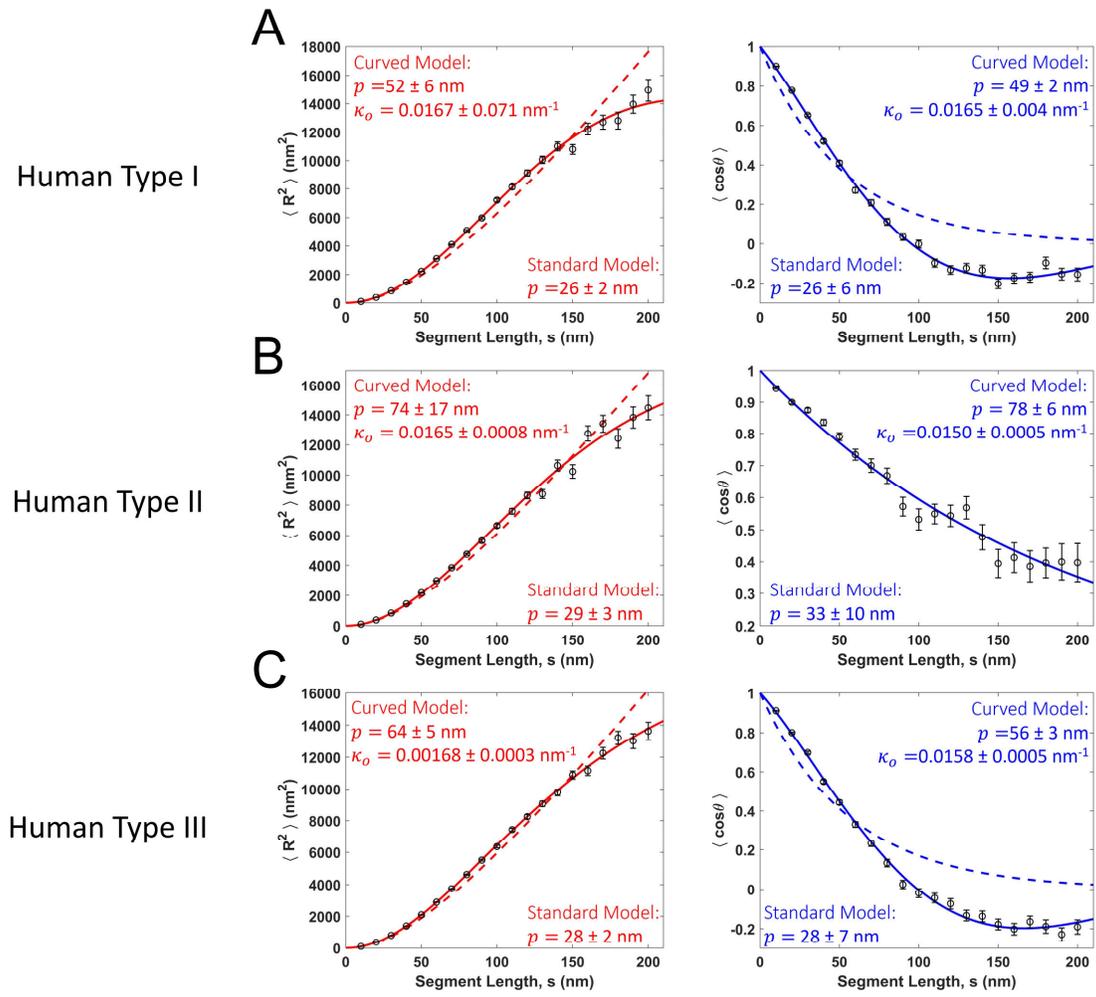

**Figure S7.** Standard (dashed lines) and curved (solid lines) WLC model fits to (A) type I, (B) type II and (C) type III human collagens deposited from 20 mM acetic acid. $\langle R^2 \rangle$ fits are shown in the left column in red, and $\langle \cos\theta \rangle$ fits are shown in the right column in blue.

**Supporting References**